# Resistive signature of excitonic coupling in an electron-hole double layer with a middle barrier

Xingjun Wu,[1,*] Wenkai Lou,[2] Kai Chang,[2]

Gerard Sullivan,[3] and Rui-Rui Du[1,4,*]

[1]*International Center for Quantum Materials, School of Physics, Peking University, Beijing 100871, China*

[2] *SKLSM, Institute of Semiconductors, Chinese Academy of Sciences, Beijing 100083, China*

[3] *Teledyne Scientific and Imaging, Thousand Oaks, California 91603, USA*

[4] *Department of Physics and Astronomy, Rice University, Houston, Texas 77251-1892, USA*

*Email address: wxj_icqm@pku.edu.cn

rrd@rice.edu

***ABSTRACT***

We study the interlayer scattering mediated by long-range Coulomb interaction between electrons (density *n*) and holes (*p*) in a double-layer system. The gated device is made of InAs (*e*) and InGaSb (*h*) quantum wells separated by a AlSb middle barrier such that the interlayer tunneling is negligibly small. By using independent-layer contacts we measure the transport tensor $\rho_{xx}$ and $\rho_{xy}$ that are solely from the InAs layer, while sweeping *p* in the InGaSb layer. We found a strongly enhanced resistive scattering signal as the carrier densities approach a total charge neutrality, *n* = *p*, which cannot be described by the Fermi-liquid theory. Results of data analysis for density, temperature, and magnetic field dependences are consistent with the emergence of excitonic coupling between the two layers, stressing the dominance of Coulomb interaction even in the presence of disorder.

*Introduction-* Interlayer interaction in closely spaced two-dimensional systems can produce a wealth of exotic phenomena with no counterpart in the single-layer case [1-10]. For example, even denominator fractional quantum Hall states at the total filling factors $\upsilon=1/4$ and $1/2$ are discovered in high mobility GaAs double-layer or wide quantum -well (QW) structures [1-4]; conversely, quantum Hall states at selected-integer filling factors in double –quantum -well (DQW) structures are found missing in high –magnetic -field regime [5,6]; and a pinned bilayer quantum Wigner crystal is observed originating from the intralayer and interlayer interactions [7]. Excitonic interaction in double-layer systems, as one of the most interesting examples of interlayer interactions, has produced a large number of many-body phenomena. Exciton condensation in quantum Hall (QH) bilayer GaAs system at total filling factor $\upsilon=1$ [8-14] has been well studied and recently extended to graphene double layers [15,16]. Enhanced resonant interlayer tunneling [8], dissipation-less counter-flow transport [9-12] and perfect Coulomb drag signals [14,15] are the manifestations of interlayer excitonic correlation. Moreover, in coupled electron-hole double layers, an electron confined in one layer may bind with a hole in the other layer, spontaneously forming excitonic states in zero magnetic field [8-16]. The possibility of realizing exciton condensation in such a novel setting has been discussed in a large number of theoretical works [17-25]. The recent observation of strongly enhanced tunneling at total charge neutrality in double-bilayer graphene-$WSe_2$ shows encouraging progress in realizing the spontaneous exciton condensate in zero magnetic field [26].

Due to a unique inverted band structure with a finite overlap of the InAs conduction band and GaSb valence band, electrons and holes could coexist in InAs/GaSb QWs, but partially confined in the respective InAs or GaSb QW. Spontaneous formation of an excitonic ground state in zero magnetic field in this material system has been proposed in Refs. [20,21]. Recently, the evidences for a BCS-like gap were observed in both terahertz transmission and electrical transport measurements, in an inverted, dilute InAs/GaSb system [27], consistent with a topological excitonic insulator. In this paper, we investigate an electron-hole double-layer with a middle barrier, in which electrons and holes of individual layers are decoupled electrically. In such structure the evidence for excitonic coupling by Coulomb interaction can be studied without concern for the role of interlayer tunneling.

*Devices and measurements-* The sample used in our experiment was grown by molecular-beam epitaxy on (100) InAs substrate, consisting of 9.5nm InAs and 5nm $In_{0.25}Ga_{0.75}Sb$ layers separated by a 10nm-thickness AlSb middle barrier. Here the GaSb layer is replaced by a ternary compound $In_{0.25}Ga_{0.75}Sb$, which is about 1 % compressive strain [28]. As a result the hole carriers should acquire a reduced mass in the plane, making a better match with the electron mass in InAs. The DQW structure is encapsulated between two AlGaSb barrier layers. The sample was wet etched into a standard Hall bar of 20×40−μm size. Ohmic contacts for connecting electron layer were made by etching selectively down to the InAs QW and depositing Ti/Ni/Au without annealing, while Ohmic contact for connecting both layers was made by depositing Indium with annealing. Dual gates were fabricated on both sides with aluminum. While the InAs substrate is conductive at low temperature and can serve as a global bottom gate, often it is desirable to make a local gate. To this end we adopt the "flip-chip" process [29] where the front side of the structure bonds to a supporting GaAs substrate, and the InAs substrate is then selectively etched, leaving a ~0.2-μm-thick DQW and the buffer layer. A metallic layer is subsequently deposited onto the buffer layer surface and patterned lithographically into a local bottom gate [Fig. 1(a)]. Details can be found in Supplemental Material [30].

The electron (hole) density at zero gate bias is $11 \times 10^{11} cm^{-2}$ ($7 \times 10^{11} cm^{-2}$), the low-temperature (~0.3K) mobility of the electron layer is ~ 10 m$^2$ V$^{-1}$s$^{-1}$, and the hole layer shows a lower mobility of ~ 1 m² V⁻¹s⁻¹ partly due to strain. Overall either the electron or hole mobility is lower than that of InAs/GaSb, attributed to the increasing scattering from interface roughness with AlSb, and to the fact that AlSb is susceptible to oxidation during processing. As a result, the present samples should be in the weakly disordered regime as far as single-electron transport is concerned. Nevertheless, as we shall elaborate throughout the paper, the Coulomb interactions are still dominating over the disorder effect, such as Anderson localization, in our electron-hole double-layers.

Fig. 1(c) shows the schematic view of our measurements. The bottom hole layer was gated by a voltage $V_{BG}$, while the top electron layer was gated by a voltage $V_{TG}$. Two layers were connected with each other through a common grounding point. In this measurement, we applied a small alternating current (~100nA) into the higher mobility InAs layer and studied this top layer using a low frequency lock-in setup through measuring longitudinal and Hall resistivities, $\rho_{xx}$ and $\rho_{xy}$. By measuring $\rho_{xy}$ in magneto-transport, we can easily determine $n$ in the top layer. In this configuration, we keep a series of nearly constant hole densities $p$ in the bottom layer with $V_{BG}$ and sweep $V_{TG}$ to vary $n$ in the top layer. Ideally, due to the effective screening of the inserted hole layer, $V_{BG}$ would not vary $n$ in the top layer unless $p$ in the bottom layer is extremely low. Otherwise, the electron-layer longitudinal resistivity $\rho_{xx}$ in this configuration can be derived phenomenologically from the Drude-like formula [31]:

$$\rho_{xx} = \frac{m_1}{e^2 n_1}\left(\frac{1}{\tau} + \frac{1}{\tau_D}\right) \quad (1)$$

Where $m_1$ and $n_1$ are the electron effective mass and density in the top layer, and $\tau^{-1}$ and $\tau_D^{-1}$ are the scattering rates from intralayer and interlayer contributions. Due to the effective screening of the hole layer, $V_{BG}$ would not change the intralayer scattering rate $\tau^{-1}$ in the top

`4`

layer. Thus, the $V_{BG}$-dependent $\rho_{xx}$ is a good approach to study the interlayer scattering rate $\tau_D^{-1}$ analogous to the drag resistivity of Coulomb drag measurement [32]. Due to the technical problem of making an independent hole contact [33] in the system, Coulomb drag measurement seems to be difficult. In spite of this, the current measurement circuit, in fact, presents the effect of Coulomb drag between the two layers in terms of interlayer scattering, where the hole layer is in an open-circuit.

Fig. 1(b) shows the corresponding result of eight-band *k·p* calculation. It has been known that the interlayer tunneling in InAs/GaSb DQWs would lead to a hybridization gap opening at the crossing points between the electron and hole dispersion relations [34]. By inserting a 10-nm-thick AlSb barrier, we find that the tunneling-induced hybridization gap cannot be resolved in our numerical calculations, consistent with previous experiment reports [35,36](*i.e.* the critical AlSb thickness for a two-independent-layers system is 2 nm). Experimentally, by independently contacting the InAs layer, single-layer characteristics have been observed in magneto-transport measurements (see supplemental materials). The interlayer tunneling resistance values are estimated on the order of at least tens of megaohms. The single-layer resistance measurement errors associated with finite interlayer resistance shown in this paper are on the order of less than 1%.

Figure 2 shows the gate-voltage dependent $\rho_{xx}$ and *n* in the InAs layer. Here *n* is determined from the low-field slope of Hall resistances and $V_{BG}$ is chosen at certain voltage values where, as we will see later, the holes screen effectively to the back gate potential. Hole densities *p* at $V_{BG}$=2.5, 5, 7.5V in the figure correspond to 4.9, 2.4, and $0.4\pm0.1 \times 10^{11} cm^{-2}$ respectively, which is determined from a two-layer-connected sample of the same wafer. From the top panel in Fig. 2, it can be found that, $\rho_{xx}$ increases rapidly first as the electrons in the top layer are being depopulated. When $V_{TG}$ exceeds -1.6V, the increasing rate of $\rho_{xx}$ becomes much slower. The top gate could not tune the electron Fermi level effectively any more. Meanwhile, we observe the gate hysteresis of $\rho_{xx}$ on opposite-directed sweeps of $V_{TG}$ (not shown). This is likely to be caused by some disorder-induced charge donor states around this energy level, *e.g.*, due to the surface donors [37], deep donors in AlSb [38] or interface donor states at InAs/AlSb [39]. Such donor states would trap and donate electrons in response to the increase and decrease of $V_{TG}$, which impedes $V_{TG}$ from tuning the Fermi level effectively. By applying a more positive $V_{BG}$, we find $\rho_{xx}$ increases further and *n* in the top layer can be tuned lower. This could be due to the band bending caused by increasing $V_{BG}$. By increasing the perpendicular electric field across the sample, the band bending in the InAs electron well (which is relatively wide) would result in an upward shift of the conduction band edge at the center of the InAs well, relative to the donor levels like those at the InAs/AlSb interface. Thus the electron density could be tuned lower owning simply to a reduction of donor states above the electron Fermi level. We would also expect an obvious change in the electron density when the electron chemical potential in InAs is tuned to cross the center of the energy levels of the donor states, which is observed around $V_{BG}$ =6V in Fig. 3(b).

*Enhanced interlayer scattering around charge neutrality in zero magnetic field-* In order to analyze the interlayer Coulomb interaction further, we map the longitudinal resistivity $\rho_{xx}$ symmetrically as a function of $V_{TG}$ and $V_{BG}$ at $B = 0T$, as shown in Fig. 3(a). The three cases of $V_{TG}$ = -1.98, -1.72, -1V are extracted and plotted in Fig. 3(b), 3(c), 3(d), respectively. From Fig. 3(d), it can be found that, due to the effective screening of the hole layer, *n* keeps a nearly constant value of $4.8\times 10^{11} cm^{-2}$ on sweeps of $V_{BG}$ ($V_{BG} < 8V$) and $\rho_{xx}$ remains invariant as well. When $V_{BG}$ exceeds 8V, $p$ in the bottom layer has become very low, and it can only screen a fraction of the electro-static potential from $V_{BG}$, resulting in $n$ in the top layer increasing and $\rho_{xx}$ subsequently decreasing. The partial screening effect can also be found in Fig. 3(b) and 3(c). As $V_{TG}$ is applied more negatively, the Fermi level moves downward into an energy range occupied by charge donor states, and then is nearly pinned around this energy level (this level is located around $n = 1.5\times 10^{11} cm^{-2}$, seen in Fig. 3). By increasing the perpendicular electric field with a more positive $V_{BG}$, the band bending in the InAs well would result in a relative shift between the conduction band edge at the center of the InAs well and the donor levels like those at the InAs/AlSb interface, and thus $n$ could be tuned lower as discussed in Fig. 2.

The most interesting observation is that $\rho_{xx}$ dramatically displays a strongly enhanced signal on sweeps of $V_{BG}$ in the regime of $V_{TG} < -1.7V$ ($n < 1.2\times 10^{11} cm^{-2}$) as shown in Fig. 3(a). When $V_{TG} = -1.98V$, the signal increases by almost 50% and is localized around total charge neutrality ($n \sim p = 0.8\pm0.15 \times 10^{11} cm^{-2}$). This is our main finding, which we will discuss below. According to Eq. (1), $\rho_{xx}$ depends on two relevant parameters -electron density $n$ and low-temperature total scattering rate $\tau_t^{-1}$. It can be seen clearly from Fig. 3(b), the evolution of $\rho_{xx}$ roughly follows $n$ except around the signal position, where $n$ continues to vary smoothly. The rapid rise of $\rho_{xx}$, or equivalently, a sudden jump of the scattering rate around total charge neutrality, is strongly suggestive of the formation of excitonic coupling between the two layers. The Fermi-liquid theory dealing with interlayer scattering seems to be invalid in this case [31]. In Fermi-liquid theory, interlayer Coulomb interaction is usually assumed to be weak, and the transport properties of each layer are often expected to be dominated by disorder, *i.e.*, $\tau \ll \tau_D$. These assumptions apply to most cases. However, if this is the case then $\rho_{xx}$ in Eq. (1) should have difficulty reflecting interlayer Coulomb scattering; thus, our observation seems nontrivial. To be specific, the Fermi-liquid result is obtained under this assumption: $\ell \ll d$ [31], where $\ell$ is the electron mean-free path and $d$ is the interlayer separation. In our case, $\ell$ at $V_{TG}$ = -1.98V can be derived quantitatively by fitting Eq. (1), $\ell$ = 15nm(see supplemental materials), comparable to $d$ = 10nm, which motivates us to think about other scattering mechanisms. Another possible explanation is based on the phonon-mediated interaction, and this can be ruled out by temperature dependence (shown in Fig.4). Interlayer scattering assisted by phonons is expected to be enhanced as *T* increases, in contrast with our observation that the scattering signal is submerged with *T* increasing.

It is illustrative to consider the inter-particle distance $r_d$ in the top layer around the peak position, $r_d = n^{-1/2}$ =35nm at $V_{TG}$ = -1.98V, which is larger than interlayer distance ($d$ =10nm). Excitonic coupling between two layers make it possible to form a collective state like the quantum Hall bilayer state [13]. As the concentration of each carrier type decreases, interlayer excitonic correlation is expected to increase due to the reduced screening. This can account for the exacerbated interlayer scattering signal with lower *n* in Fig. 3.

*Discussion of the origins of the strongly enhanced scattering signal-* To gain further insight into the origin of the enhanced scattering signal around $n \sim p$, magnetic-field and temperature dependences are measured. Figure 4(a) shows the temperature dependence. We can find a decreasing resistivity background with *T* increasing. This could be due to more donor states becoming active to release electrons to participate in the conductivity at higher *T* [40]. In addition to the decreasing resistivity background, it is interesting to find that, the resistivity signal becomes weakened as *T* increases, and finally is submerged by the background around 15 -19 K. This observation contradicts with the Coulomb scattering phase-space argument [31], that is, increasing *T* will increase scattering phase space; thus, an increasing resistivity is expected. It implies a character of an insulating state instead. This insulating state characterized by one layer is difficult to understand in terms of a picture of two-uncorrelated-layer structure. We explain this unusual insulating state as a collective behavior of a double-layer system probably due to the opening of an interlayer exciton-like gap.

Parallel magnetic field dependent $\rho_{xx}$ also has been presented in Fig. 4(b). It can rule out the possibility that this collective insulator state arises from an interlayer tunneling-induced gap. With applying $B_{//}$, the electron and hole dispersion relations would induce a relative shift proportional to their relative displacement in real space $\Delta z$. The regions where electron-hole band mixing exists are expected to move away from the Fermi energy, and thus the resistivity would decrease with increasing $B_{//}$ [41]. This is opposite to our observation. One possible explanation is that a virtual transition between electron and hole bands would be attenuated with increasing $B_{//}$, immune to an increased dielectric constant, and hence lowered exciton binding energy. Thus, the resistivity signal is expected to be strengthened with increasing $B_{//}$ as shown in Fig. 4(b). We further examine the perpendicular magnetic field dependent $\rho_{xx}$ at $V_{TG}$ = -1.98V as shown in Fig. 4(c). Interestingly, we find the signal shows a significant $B_\perp$ dependence, where a ~$B^2$ dependence is observed as shown in the inset. An enlarged scattering phase space due to the increase of Landau level degeneracy at higher fields cannot explain the strong field dependence. In the assumption of a constant interlayer perturbation, the scattering rate based on Fermi's "golden rule" is expected to grow linearly with $B_\perp$, slower than the quadratic dependence, indicating a $B_\perp$-promoted interlayer interaction. This is consistent with the prediction of interlayer exciton-like interaction. $B_\perp$ is expected to enhance the confinement and, hence, the binding energy.

In conclusion, we have measured longitudinal- and Hall signals of a InAs layer in the presence of a spatially close InGaSb layer. We observed a strongly enhanced interlayer scattering around charge neutrality. Results of data analysis for density, temperature, and magnetic field dependences are consistent with the emergence of excitonic coupling between the two layers, stressing the dominance of Coulomb interaction even in the presence of disorder.

*Acknowledgements.* The work at Peking University was financially supported by National Key R and D Program of China (2017YFA0303301). The work at Rice University was funded by NSF Grant No. DMR-1508644 and Welch Foundation Grant No. C-1682. WKL and KC were supported by NSFC (Grant No. 11434010).

# Figure Captions

**FIG. 1.** (Color Online) (a) Sketch of InAs/InGaSb double layer device. Electrons and holes are confined in InAs and InGaSb QWs respectively. (b) Eight-band $k \cdot p$ model result of the band structure for InAs/InGaSb double layer. (c) Schematic view of our device and measurement geometry. The device has a 10nm AlSb middle barrier.

**FIG. 2.** (Color Online) T=0.3K, $\rho_{xx}$ and corresponding electron density $n$ in the electron layer as a function of $V_{TG}$ for different bottom gate biases stepped in units of 2.5V.

**FIG. 3.** (Color Online) (a) $\rho_{xx}$ as a function of $V_{TG}$ and $V_{BG}$ in zero magnetic field at T = 0.3K. (b, c, d) $V_{BG}$ dependent $\rho_{xx}$ at $V_{TG}$ = -1.98, -1.72 and -1 V respectively. The corresponding plotted data $n$ in top layer are determined from the small field slope of the Hall resistance.

**FIG. 4.** (Color Online) Magnetic-field and temperature dependences of $\rho_{xx}$ at $V_{TG}$ =-1.98V. (a) Temperature dependence of $\rho_{xx}$ at zero B. The resistivity signals in the rectangle are highlighted. (b, c) T = 0.3K, $\rho_{xx}$ as a function of $V_{BG}$ at different $B_{//}$ and $B_{\perp}$. The $B_{\perp}$-dependent resistivity values of the signals are plotted in the inset of panel (c). Circles depict experimental data, and solid curve depicts the quadratic fit of the data.

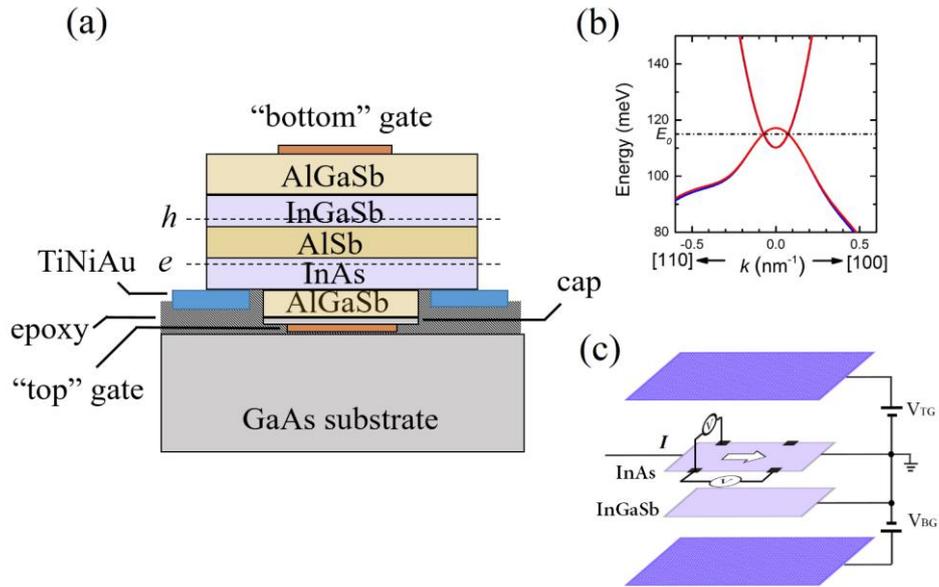

**Figure 1**

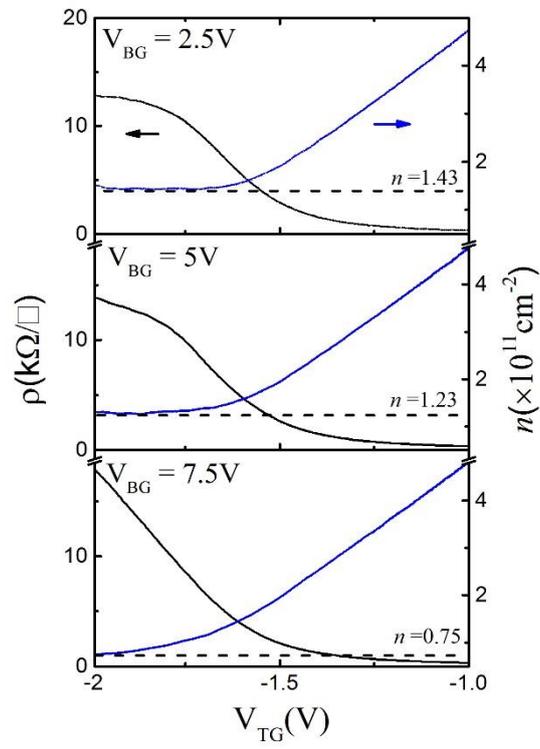

**Figure 2**

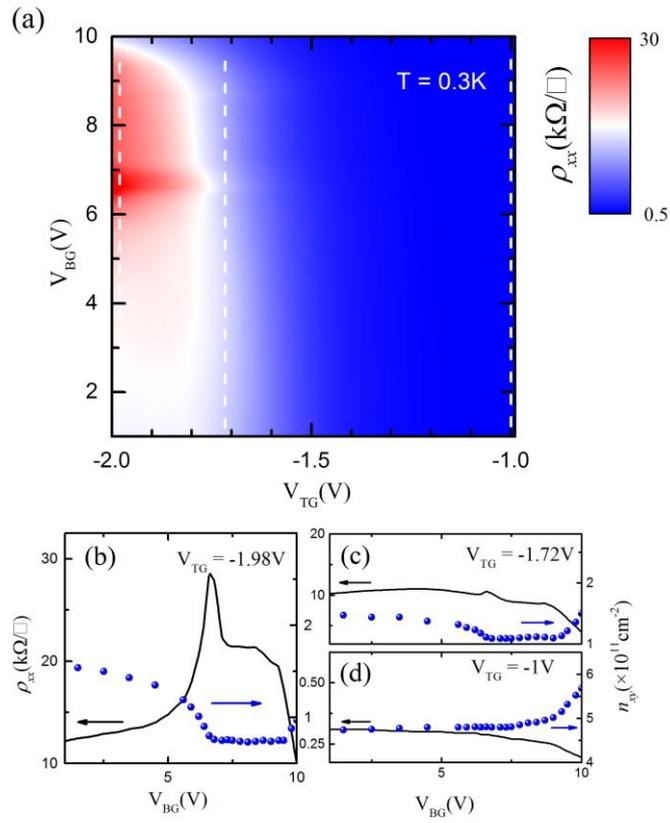

**Figure 3**

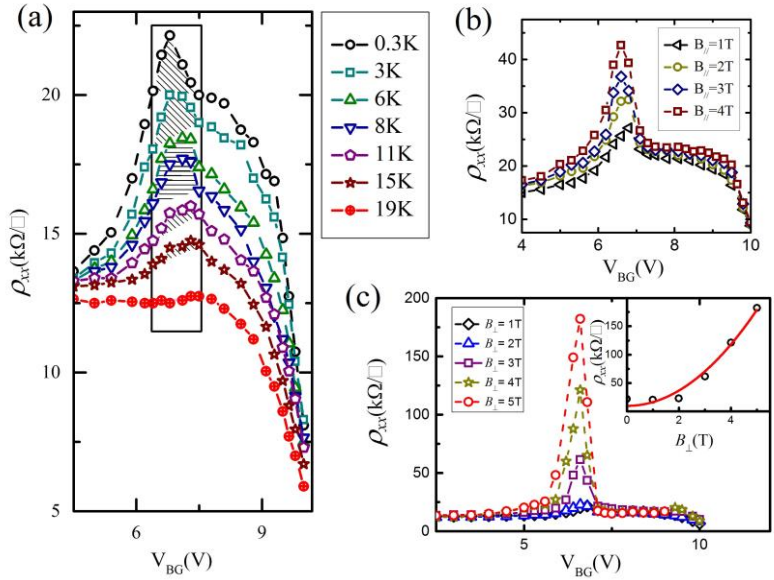

**Figure 4**